# Arthroscopic Multi-Spectral Scene Segmentation Using Deep Learning

*Shahnewaz Ali, Dr. Yaqub Jonmohamadi, Yu Takeda, Jonathan Roberts, Ross Crawford, Cameron Brown, Dr. Ajay K. Pandey*

Robotics and Autonomous Systems, School of Electrical Engineering and Computer Science, Queensland University of Technology, Gardens Point, Brisbane, QLD 4001, AUSTRALIA.

*Abstract*— **Knee arthroscopy is a minimally invasive surgical (MIS) procedure which is performed to treat knee-joint ailment. Lack of visual information of the surgical site obtained from miniaturized cameras make this surgical procedure more complex. Knee cavity is a very confined space; therefore, surgical scenes are captured at close proximity. Insignificant context of knee atlas often makes them unrecognizable as a consequence unintentional tissue damage often occurred and shows a long learning curve to train new surgeons. Automatic context awareness through labeling of the surgical site can be an alternative to mitigate these drawbacks. However, from the previous studies, it is confirmed that the surgical site exhibits several limitations, among others, lack of discriminative contextual information such as texture and features which drastically limits this vision task. Additionally, poor imaging conditions and lack of accurate ground-truth labels are also limiting the accuracy. To mitigate these limitations of knee arthroscopy, in this work we proposed a scene segmentation method that successfully segments multi structures.**

## I. Introduction

Arthroscopy, after its long clinical validation has become a widely used medical procedure to treat bone-joint ailments. During this minimally invasive surgical (MIS) procedure a miniaturized camera and tools are introduced into the knee cavity through small incisions. It offers several benefits to patients such as less surgical trauma, minimum tissue and blood loss, and quick recovery time. However, indirect vision of the surgical site, lack of perception and haptic feedback, and limited access to the operating space are the drawbacks of this surgical procedure [1]. Furthermore, unintentional tissue damage can occur due to lack of contextual information. Arthroscopy performed in an underwater environment to create additional space for tools and cameras. Imaging devices are placed at close proximity to tissue structure (about 10mm) due to lack of space and to avoid obstacles like unstructured fat tissue. Additionally, conventional imaging devices used in knee arthroscopy can support only 30-70-degree field-of-view (FoV). With these limitations only a small part of the surgical space is accessible with very limited contextual scene details. Furthermore, in an underwater environment, light diffuses into water and causes color attenuation. In this perturb environment caused by the dissolved materials like body fluid and floating tissue subsequently exhibits noisy observations. Not only that, the lighting conditions and several motions influenced by the environment and unsteady camera maneuver cause saturated and blurred frames. Therefore, most of the video frames generally exhibit insignificant information for clinical decisions and surgeons have to maneuver cameras from known structures. This additional maneuver can cause unintentional tissue damage. To mitigate these drawbacks, there is a clear demand for enhanced MIS intra-operative vision. [2] method provides a way to adjust scene illumination conditions automatically to mitigate the effect of low and excessive lighting conditions. Among others, understanding arthroscopic surgical scenes from video sequences has a long-standing demand to obtain contextual awareness [1]. In progress of surgical scene segmentation, a significant progress has been achieved in laparoscopy, however, in arthroscopy the achievements are still limited considering the limitations discussed above. The spectral reflectance is unique to tissue structure and can be a potential metric to achieve more accurate segmented map instead of pixel level information such as texture and features. When light interacts to a surface particle, the energy loss occurs at different bands depending on the composition of the interacting material. It has several diagnostic and therapeutic applications [3-4] under a precisely controlled lighting and experimental conditions. In previous study, it has been found that at dynamic lighting conditions the pose of the camera and light source limits the assessment of multispectral response to identify tissue type under diffuse reflection. The main reason is that the single reflectance gets contaminated by the contributions of other reflectances in the confined knee space. Moreover, some tissue such as dense collagen produces almost similar reflectance like Anterior Cruciate Ligament (ACL). Additionally, scattering due to dissolved body fluids, debris and color attenuation are the common drawbacks that can affect single multi-spectral based methods. In order to overcome the limitations of single reflectance-based method, in this work entire multi-spectral frames are used in a deep neural network to achieve multi class segmented map for arthroscopic video frames.

## II. Background Study

To alleviate the requirements of necessary hardware and lighting conditions, recently reconstruction of multi spectral image from RGB pixel value is gaining a lot of research attention by the computer vision community. Previously, it has been studied by the computer graphics community and widely used in bio-photonics to diagnose tissue structure in a very controlled environment. Wieiner et. al proposed a method to reconstruct spectral response capture by a camera from its filter's response namely RGB [5] which defined through the camera response function as follows;

$$P_{i<R,G,B>}(x, y) = \int t_{<R,G,B>}(\lambda) E(\lambda) S(\lambda) r(x, y: \lambda) d\lambda \quad (1)$$

Here, $t_{<R,G,B>}(\lambda), P_{i<R,G,B>}(x, y), E(\lambda), S(\lambda), r(x, y: \lambda)$ stand for permeability of color filters, pixel in RGB space, spectrum of the illuminant, sensitivity of the camera and reflectance spectrum in spatial domain respectively. In his method camera function is estimated and resultant transformation matrix transfer pixel value to its corresponding reflectance for known illuminant. After that there are several improvements are achieved [6,7] including recently proposed deep learning-based methods that predict multi and hyper spectral images from RGB value. In this work, the spectral reflectances are obtained from otsu et.al. [8].

## III. Multi-structure Segmentation

In order to address this issue, an experiment is carried out considering a patch of a size of $nxm$ of reflectances unlikely to single reflectance. The idea is to find out the discriminative features of the patches, considering that some of the pixels in a patch can get contaminated. A set of frames are selected from different arthroscopic videos where camera position and the surface are not directional, and the observing surface is slanted. These conditions reduce the chance of specular reflectance capture by the camera. Ideally, if the patches are not overlapped between the anatomical structures, then it will express different structure robustly even if some of the pixels get contaminated inside a patch.

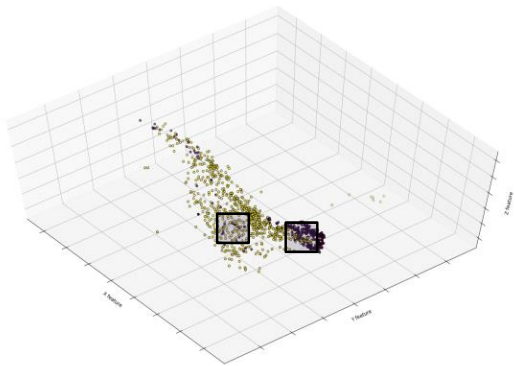

(a)

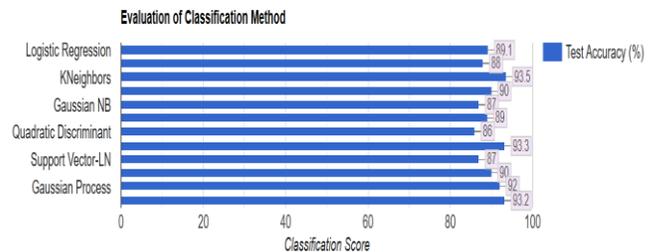

(b)

Fig.1. Figure represents the patches obtained from patch-dataset constructed from relatively high quality of images. PCA is used to reduce the dimension into three in order to plot the dataset, therefore X,Y, and Z axes are representing PCA features. Patches with window size 7x7 do not uniquely represents binary class, there are potential overlaps are observed in the dataset even the images are free from poor imaging quality.

The experimental result of patch-based approach for binary segmentation is summarized in Fig.1. Patches are extracted from each frame and then they are stored along with their labels. The size of each patch is 7x7x36 where 36 represents the spectral channels of each pixel that corresponds to the reflectance from 380nm to 730 nm. Principal component analysis (PCA) is applied to get patches into 3-dimensions that correspond to three PCA features in order to plot random samples from patch-dataset. The resultant patches are presented in Fig.1. Moreover, a set of machine learning algorithms is applied on the dataset and corresponding scores are represented in the Fig.1-b.

In concussion, it is confirmed that the patch of size 7x7 does not convey robust discriminative information for two class segmentation. Overlapped patches has been observed between the tissue structures caused by the following basic factors; i.) the close proximity between the camera and different tissue structures, ii.) close proximity of different tissue structures, iii.) curved and multiple slanted surfaces. However, patch-based method has its own shortcomings. Each pixel corresponds to a patch of 7x7 window size; therefore, patch extraction and classification process increases computational time for each video frame thus it decreases system performance. Moreover, selection of an appropriate patch size is a challenging task, larger window may increase the segmentation accuracy in some situations but decreases computational efficiency.

In progress of semantic segmentation, considering the above findings, the further open hypothesis is constructed as follows, local object contextual features and scene details such as context arrangements, shape etc. can further improve the semantic segmentation. Here, local object context stands for a set of features of each tissue structure in a frame at different light wavelengths, a multispectral spatial feature ranging from 380nm to 730nm. Scene details are the set of information in a frame such as their structural information like shape and their arrangements.

To these contexts, UNet – a deep learning architecture and its variants provide a strong benefit to meet these above requirements. The other strong ground for UNet architecture is that its architecture can work with limited training samples. It is of great influence on the learning strategy of biomedical images due to it supports the above domain specific requirements.

The classical UNet architecture has two distinct paths to learn about the context of an image. In UNet architecture, as stated in the Fig.2, the contracting path also known as down sampling path, learns the context of the image in spatial domain where the up sampling path also known as an expansive path, learns the precise location of each scene details. Moreover, skip connections between up sampling and down sampling paths preserve the spatial information between each contracting stage.

Recently, researchers made some progress in knee segmentation [9-15] using multimodal images in a UNet architecture. In this progress, some researchers used Ultrasound and magnetic resonance images to segment knee cartilage in a deep learning framework using UNet. In their proposed method, Maria et.al used ultrasound images to segment femoral cartilage [9]. In their method, the UNet is trained with an average of 15,531 training and 3,124 testing data.

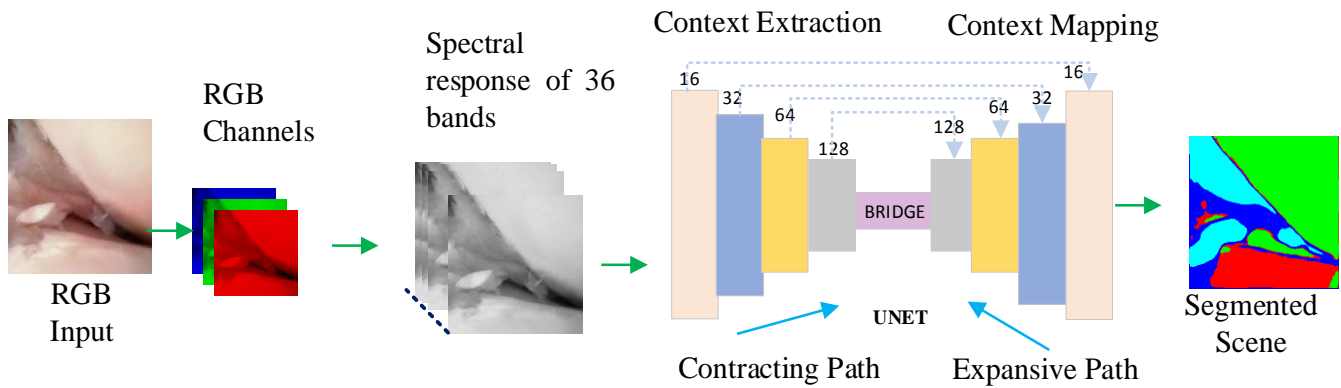

Fig.2: System overview of the proposed method; multi-structure tissue segmentation of knee arthroscopy using surface reflectance. Here, the surface reflectance for each pixel is calculated from RGB response that produces an image with spectral reflectance of 36 bands. The image is then feed to the deep learning network, UNet. Contracting path of UNet uses dilated convolution and subsequently extracts spatial feature from these 36 multi-spectral bands using relatively wider FOV while the total trainable parameter remains same. The extracted feature then labelled and mapped to the final segmentation map.

In progress towards semantic segmentation of knee arthroscopic scene, our medical robotics group at Queensland University of Technology (QUT) has significant progress [1,12]. Among others, Y. J. et al. in their work used RGB images obtained from arthroscope [1]. According to his article Y. J. et al. achieved automatic scene segmentation from arthroscopic images. In their work, RGB video frames are used to train the deep learning network UNet and its variant UNet++ to achieve multi structure knee arthroscopic instance segmentation.

Considering the benefits, in this work UNet architecture is revisited in a multispectral domain. Each RGB image is transformed to its spectral reflectance domain following the previously stated method. So, the input channels for UNet implementation is set to 36 where conventional implementation uses one or three channels of data. Therefore, multispectral band precise spatial features are extracted by the UNet model when convolution is performed.

The UNet down sampling path is built using the five blocks of the contraction layer. Each contraction layer contains two successive convolution layers with kernel size of 3x3 and one rectified linear activation function (ReLU) and then spatial context map is down sampled by max pooling operation with pool size 2x2. In order to prevent network overfitting a dropout factor of 0.1 is used. Padding is used to get the same resolution of input and output image. Kernel initializer is used to set initial weights of the convolutional layer. Adjacent pixels have the highest priority to belong to the same reflectance, either specular or diffuse. Dilation rate is set to 2, so that it can avoid adjacent pixels having a wider field of view while the number of parameters remain the same. Softmax activation function is used at the final layer. Both Adam and Stochastic gradient descent (SGD) optimizer is evaluated to train the model separately. In this implementation SGD outperforms to Adam and hence the SGD is used with learning rate 0.001. This UNet implementation used categorical cross-entropy (CE) loss function;

$$CE = -\sum_{i}^{C} t_i \log t_p \qquad (1)$$

Here, $C$ denotes the total category that the class number, $t_i$ is the ground-truth label, $t_p$ is the predicted label that is the outcome of the softmax activation function.

In most video sequences, femur occupies most of the frame area where meniscus and others exist in a small image area. Moreover, often clinicians use femur as landmark for the knee anatomy, the presence of femur bone is relatively high as a consequence our arthroscopic video sequences are experiencing highly imbalanced data. Class imbalanced data are address through computation of class weight. It is worth noting that, the rotation of camera and leg manipulation can produce frames of the same context which can have different visual appearances. UNet provides a way to learn precise location of the context through expansion path. To capture scene context at different locations, data augmentation has been performed. In this augmentation phase each training sample is multiplied by its following augmented version.

## IV. RESULT

Fig.3 represents the sample result of our method validated on the arthroscopy video frames of three cadaver knee samples. Insertion of union (IoU) is used to measure segmentation accuracy. Analytical information reveals that the bone-cartilage received highest segmentation accuracy among the others. In this work, test is performed on selected images. Total 400 high quality images are selected which depict less noises, debris and blur effect. The achieved segmentation accuracy for bone is 92%. The segmentation accuracies for the tissue structures ACL, meniscus are 74 and 61 respectively. Table I represent overall test accuracy performed on three cadaver datasets. It is worthful to note that at this stage, only last cadaver dataset contains relatively high quality of images but the geometrical structure are different due to tissue degeneration.

| Test Data | Bone | ACL  | Meniscus |
|-----------|------|------|----------|
| Set. 1    | 0.81 | 0.52 | 0.401    |
| Set. 2    | 0.78 | 0.43 | 0.36     |
| Set. 3    | 0.83 | 0.59 | 0.45     |

Table I. Result obtained from three different cadaver knee dataset. It

From this analysis, it is apparently confirmed that our multi-spectral method reveals a better solution with higher segmentation accuracy. The accuracy can be improved if more scene contextual information is provided. Due lack of good quality images and deliberately improvement of imaging system at this time the approach is limited.

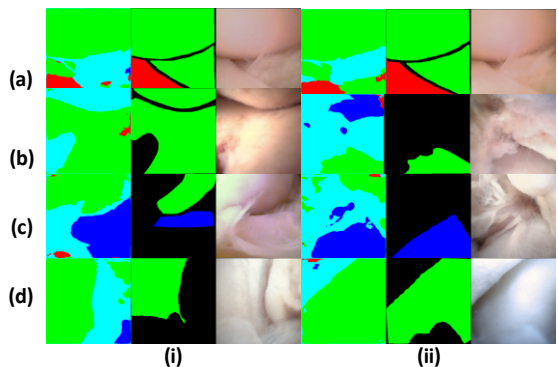

Fig.3. Figure represents the sample outcome of the proposed segmentation method. First row represents the cadaver sample where lighting source is applied to the surface through a different incision. With relatively poor imaging conditions the proposed method segments the scene with high accuracy. Second, third and fourth rows are representing the segmented surgical scene where in some context it captures more details compared to ground-truth label obtained from clinician's efforts. The outcome of the proposed method again justified by the clinician in order to validates the model accuracy. For instance, in the image (c)-(i) and (c)-(ii), the ground-truth data are not expressing a fine details of the scene but a tiny part of the tissue structure acl and bone cartilage in the first scene and only acl in the second scene. Compare to the ground-truth labels, the proposed method successfully captured more scene details.

Bone cartilage tissue received the higher segmentation accuracy respect to others. In arthroscopic video frames bone sequences are exist in most of the frames and generally, the bone cartilage especially femoral cartilage provides more details in a scene part than others. As previously mentioned in order increase the confidence, surgeons often use femur bone as a landmark, hence, it is repeatedly observed in many frames.

Fig.4 represents the sample outcome from three cadaver test samples. Fig.4-(a) are images captured by a different camera sensor than others and used an external lighting source. Apparently, these images received more color distortion, and sometimes compromised to blur and distortion (defocused and noise). The difference in color of these video sequences does not compromises the accuracy until the lighting scheme supports visible light spectrum. In those two scenes, segmentation for meniscus region (in red) is slightly compromised compared to ground-truth and those regions are in close proximity to bone-meniscus tiny cavity and includes shadow area. However, other bone area (tibia-cartilage, in cyan color) though provides an impression of false segmentation but it is confirmed by the orthopedic that on top of tibia-cartilage there are some soft-tissue (possibly synovium) and it is more accurate compare to the ground-truth data that is manually labelled by other the clinicians. In the column (i) of Fig.4-(b) apparently more informative than the ground-truth data. Clinicians only provided bone label with high confidence where the proposed method can justify the presence of other soft-tissue and the ACL tissue structure. Similarly, in the column (i) of Fig.4-(c) the ACL bundle is identified by the method more accurately than the clinician did. Among the other reasons stated above, clinicians experience more difficulty to identify tissue structure from the frame having small FOV especially when the tissue structure are slightly visible or surrounded by the unstructured tissue type such as fat. To this context our proposed method reflects an outstanding matching.

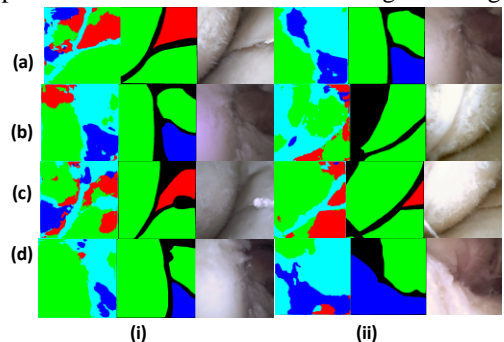

Fig.4. Figure represents the sample outcome of the proposed method having relatively lower accuracy. At this current stage, the network is only trained with the data samples having relatively high quality of images with fine details. When the proposed model is not trained with the degenerated knee anatomy, for instance the image set presented here in first row, it is still capturing the details of the scene but compromised to the degenerated cartilage section. Here the degenerated cartilage experiences some fiber like tissue on top of the femur bone that produces high frequency components on the image plane as a result of light scattering. Rough surface with high frequency components generates information close to ACL tissue bundle, hence, the method segments some area of the cartilage section as ACL. Moreover, being different than the smooth bone surface, where the density of the degenerated cartilage appears like fibers is relatively low produces a small difference that ACL and hence, it close to meniscus tissue type and segmented as a part of meniscus in some area. It is also worthwhile to mentioned that the network is trained with the spatial features at different wavelength obtained from relatively high quality of images captured in our last cadaver arthroscopy experiment. The other previous experiments have suffered from poor lighting conditions, image noise and blur. This figure confirms that the proposed method though it is providing a moderate segmentation accuracy, can slightly compromise its accuracy to the poor imaging conditions but can performs better with high quality of images.

There are significant differences between the healthy and unhealthy knee anatomy. Degenerative cartilage tissue of bone type can create a different level of complexity as depicted in Fig. 4-a. In the column (i) represents degenerated cartilage surface on the left upper corner. Roughness of the unhealthy cartilage produces light scattering, and the reflection is deviated from the smooth surface. Other type unhealthy cartilage can be observed where cartilage decayed, and bone surface appears in reddish color mostly due blood. We experienced it with one cadaver arthroscopic samples that is not used during the training.

## V. CONCLUSION

MIS procedure is performed through the indirect vision of surgical space that reinforces the demand of augmented vision to improve its accuracy. However, surgical images of MIS are highly texture-less and subsequently it is showing very limited image features. These are restricting the applicability of vision-based solutions such as surgical scene segmentation that has an obvious demand to improve quality of life. In this work, surface reflectance is evaluated as an alternate to pixel information such as texture or feature to perform scene segmentation. This research confirms that, surface reflectance can be great alternative to attain surgical scene segmentation.

In this article, we are proposing arthroscopic scene segmentation based on multi-spectral response reconstructed from RGB triplet thus it does not require any additional imaging system or optical tool. Moreover, this work confirms that the segmentation accuracy of the proposed method solely depends on the spatial features of the anatomical structure at wavelength from 380nm to 730nm irrespective to texture and features details.

In this work, deep learning framework UNet is implemented to adopt multispectral data. There are 36 of spectral bands are used. Network extracts spatial characteristics at these 36 bands by performing two-dimensional convolution in each band and subsequently learn of the location along with its label. Therefore, the network realizes the spectral response of the concern tissue types as well as the geometrical structure. Training samples contains relatively high quality of images. The high-quality frames increase the network capability to segment.

ACKNOWLEDGMENT

This work is supported by Australian Indian Strategic Research Fund (AISRF), The Medical Engineering Research Facility and QUT Centre for Robotics.

REFERENCES

[1]. Y. Jonmohamadi et al., "Automatic Segmentation of Multiple Structures in Knee Arthroscopy Using Deep Learning," in IEEE Access, vol. 8, pp. 51853-51861, 2020, doi: 10.1109/ACCESS.2020.2980025.
[2]. Shahnewaz Ali, DrYaqub Jonmohamadi, Yu Takeda, Jonathan Roberts, Ross Crawford, Cameron Brown, Ajay K. Pandey "Supervised Scene Illumination Control in Stereo Arthroscopes for Robot Assisted Minimally Invasive Surgery" submitted on IEEE Sensors Journal
[3]. de Boer, L.L., Molenkamp, B.G., Bydlon, T.M. et al., "Fat/water ratios measured with diffuse reflectance spectroscopy to detect breast tumor boundaries", Breast Cancer Res Treat, 152: 509, 2015
[4]. Blaž Cugmas et.al, "Detection of canine skin and subcutaneous tumors by visible and near-infrared diffuse reflectance spectroscopy", Journal of Biomedical Optics, Vol 20, 2015
[5]. Stigell, P., Kimiyoshi Miyata, and Markku Hauta-Kasari. "Wiener estimation method in estimating of spectral reflectance from RGB images." *Pattern Recognition and Image Analysis* 17.2 (2007): 233-242.
[6]. Khan, Zohaib Amjad, et al. "Residual Networks Based Distortion Classification and Ranking for Laparoscopic Image Quality Assessment." *2020 IEEE International Conference on Image Processing (ICIP)*. IEEE, 2020.
[7]. Xu, Peng, et al. "Self-training-based spectral image reconstruction for art paintings with multispectral imaging." Applied optics 56.30 (2017): 8461-8470.
[8]. H Otsu, M Yamamoto, T Hachisuka. , "Reproducing Spectral Reflectances From Tristimulus Colours." Computer Graphics Forum 37 (6), 370-381, 2018. 10, 2018.
[9]. Antico, Maria, et al. "Deep learning-based femoral cartilage automatic segmentation in ultrasound imaging for guidance in robotic knee arthroscopy." Ultrasound in Medicine & Biology 46.2 (2020): 422-435.
[10]. Mohabir, Justin Heeralaal. Knee cartilage segmentation of ultrasound images using convolutional neural networks and local phase enhancement. Diss. Rutgers University-School of Graduate Studies, 2020.
[11]. Kumar, Dileep, et al. "Knee articular cartilage segmentation from MR images: A review." ACM Computing Surveys (CSUR) 51.5 (2018): 1-29.
[12]. Tan, Chaowei, et al. "Collaborative Multi-agent Learning for MR Knee Articular Cartilage Segmentation." International Conference on Medical Image Computing and Computer-Assisted Intervention. Springer, Cham, 2019.
[13]. Honkanen, Miitu KM, et al. "Triple contrast CT method enables simultaneous evaluation of articular cartilage composition and segmentation." Annals of Biomedical Engineering 48.2 (2020): 556-567.
[14]. Shah, Romil F., et al. "Variation in the thickness of knee cartilage. The use of a novel machine learning algorithm for cartilage segmentation of magnetic resonance images." The Journal of arthroplasty 34.10 (2019): 2210-2215.
[15]. Myller, Katariina AH, et al. "Method for segmentation of knee articular cartilages based on contrast-enhanced CT images." Annals of biomedical engineering 46.11 (2018): 1756-1767